\documentclass[amsmath,amssymb,aps,prd,11pt,tightenlines,superscriptaddress,
nofootinbib,preprintnumbers,notitlepage]{revtex4-1}
\usepackage{multirow}
\usepackage{graphicx}
\usepackage{tabularx}
\usepackage{slashed}
\usepackage{url}
\usepackage{hyperref}
\usepackage[maxfloats=100]{morefloats}
\usepackage{placeins}
\usepackage{color,xcolor}
\usepackage{booktabs}
\usepackage{bbold}
\usepackage[normalem]{ulem}
\makeatletter

\def\l@subsubsection#1#2{}
\makeatother

\newcommand{\toolfont}[1]{\texttt{#1}}
\newcommand{\ord}{\ensuremath{\mathcal{O}}}

\newcommand{\lag}{\ensuremath{\mathcal{L}}}
\newcommand{\mat}{\ensuremath{\mathcal{M}}}

\newcommand{\bpm}{\begin{pmatrix}}
\newcommand{\epm}{\end{pmatrix}}

\newcommand{\gev}{{\ensuremath\rm GeV}}
\newcommand{\tev}{{\ensuremath\rm TeV}}

\def\slashchar#1{\setbox0=\hbox{$#1$}           
   \dimen0=\wd0                                 
   \setbox1=\hbox{/} \dimen1=\wd1               
   \ifdim\dimen0>\dimen1                        
      \rlap{\hbox to \dimen0{\hfil/\hfil}}      
      #1                                        
   \else                                        
      \rlap{\hbox to \dimen1{\hfil$#1$\hfil}}   
      /                                         
   \fi}

\AtBeginDocument{
  \heavyrulewidth=.08em
  \lightrulewidth=.05em
  \cmidrulewidth=.03em
  \belowrulesep=.65ex
  \belowbottomsep=0pt
  \aboverulesep=.4ex
  \abovetopsep=0pt
  \cmidrulesep=\doublerulesep
  \cmidrulekern=.5em
  \defaultaddspace= .5em
  \setlength{\tabcolsep}{0.5em}
}
\begin{document}
\preprint{IFT-UAM/CSIC-16-072\cr
SLAC-PUB-16779\cr
TTK-16-34}

\title{Complementarity of Resonant Scalar, Vector-Like Quark and Superpartner Searches in Elucidating New Phenomena} 

\author{Anke~Biek\"otter}
\affiliation{Institut f\"ur Theoretische Teilchenphysik und Kosmologie, RWTH Aachen, Germany}
\author{JoAnne~L.~Hewett}
\affiliation{SLAC National Accelerator Laboratory, Menlo Park 94025, CA, USA}
\author{Jong~Soo~Kim}
\affiliation{Instituto de Fi­sica Teorica UAM/CSIC, Madrid, Spain}
\author{Michael Kr\"amer}
\affiliation{Institut f\"ur Theoretische Teilchenphysik und Kosmologie, RWTH Aachen, Germany}
\author{Thomas~G.~Rizzo}
\affiliation{SLAC National Accelerator Laboratory, Menlo Park 94025, CA, USA}
\author{Krzysztof~Rolbiecki}
\affiliation{Instytut Fizyki Teoretycznej, Uniwersytet Warszawski, Warsaw, Poland}
\author{Jamie~Tattersall}
\affiliation{Institut f\"ur Theoretische Teilchenphysik und Kosmologie, RWTH Aachen, Germany}
\author{Torsten~Weber}
\affiliation{Institut f\"ur Theoretische Teilchenphysik und Kosmologie, RWTH Aachen, Germany}

\date{\today}

\begin{abstract} 
  The elucidation of the nature of new phenomena requires a multi-pronged approach to understand 
  the essential physics that underlies it. As an example, we study the simplified model containing a new scalar singlet accompanied by 
  vector-like quarks, as motivated by the recent diphoton excess at the LHC.  To be specific, we investigate three 
  models with $SU(2)_L$-doublet, vector-like quarks with Yukawa couplings to a new scalar singlet and which also couple 
  off-diagonally to corresponding Standard Model fermions of the first or third generation through the usual Higgs 
  boson. We demonstrate that three classes of searches can play important and complementary roles in constraining this model.  In particular, we find
  that missing energy searches designed for superparticle production, supply superior sensitivity for vector-like quarks than the dedicated
  new quark searches themselves.
   \end{abstract}

\maketitle
\tableofcontents

\clearpage
\section{Introduction}
\label{sec:intro}

New physics beyond the Standard Model (SM) might take many forms and, once discovered, will require a complementary 
multi-pronged attack in order to understand its essential nature. In wider context, as in the case of Dark Matter, this has 
been recognized 
for a long time \cite{Arrenberg:2013rzp}.  At the LHC, such an approach will clearly be advantageous no matter what the source of the 
new physics might be.  

In this work, we illustrate the benefits of the complementarity of LHC searches by examining the specific simplified new physics scenario of an 
additional scalar singlet
and a vector-like quark doublet \cite{Xiao:2014kba,simp_higgs}, also motivated by the possible signal of a new resonance at 750 GeV 
\cite{Aaboud:2016tru,Khachatryan:2016hje}.
To study this test case, we have combined numerous new physics search channels and analyzed their effectiveness for observing this scenario.
In the process, we have found the interesting result that in some cases, the supersymmetry missing transverse energy (MET) based searches perform
better in searching for vector-like quarks (VLQs) than the specifically VLQ designed search channels.  
This provides an explicit example of the interdependency of LHC
new physics searches and shows that combining, or recasting, search results would produce improved constraints.


To be specific in defining our scenario, we set the mass of the scalar singlet to be 750 GeV and  investigate the possibility that the production and decay 
of this state is mediated by color-triplet, vector-like quarks
\cite{Benbrik:2015fyz,Chao:2015ttq,Angelescu:2015uiz,Dev:2015vjd}. More specifically, we will investigate adding a VLQ $SU(2)_L$ doublet 
whose mass is generated through the vev of the new scalar and which also has Yukawa couplings to the SM fermions via the 
usual Higgs doublet \cite{Aguilar-Saavedra:2013qpa,Buchkremer:2013bha}. In this case, there are three possible charge 
assignments for the new VLQ fermion states and in order to be concrete, we investigate each of these individually to see if they can reproduce the 
apparent diphoton production rate. We examine the branching fractions for all relevant possible 
final states of the scalar resonance in order to ensure that a given model does not produce another signal that would have 
already been seen.

The presence of the VLQs in this simplified 
model implies that they can also be produced and searched for directly at the LHC. We consider the possibility 
that these new states either predominately mix with the first or third generation SM fermions of the same charge. We then 
apply the existing ATLAS VLQ search at 13~TeV \cite{ATLAS13VLQ} along with a number of ATLAS and CMS supersymmetry searches 
at both 8 and 13~TeV that are also found to be sensitive to VLQ direct production. 

The results obtained in this paper go far beyond those from Ref.~\cite{Barducci:2014gna} where only three 8 TeV searches 
were included. Using the variety of the most recent searches, including seven at $\sqrt{s} = 13$~TeV (for a total of 105 signal regions), significantly 
improves the sensitivity to direct VLQ production and decay. In a very recent analysis, Ref.~\cite{Kraml:2016eti}, it was 
also shown that SUSY searches can constrain VLQ models which also contain dark matter particles. With our updated suite 
of searches we arrive at a similar conclusion in a set of models without dark matter in the final state. In fact, we find 
that the SUSY searches provide better sensitivity than the dedicated analysis for VLQs in both cases 
where these new states mix dominantly (but weakly) with either the first or third generation. This is 
not surprising for first generation mixing since the existing dedicated VLQ search at 13~TeV does not consider
this possibility. However the better performance of the supersymmetric search to third generation
mixing does motivate the development of improved VLQ searches that make use of missing energy\footnote{We note 
that in the finishing stages of this study, ATLAS released a new search with signal regions sensitive to VLQ's that include
missing energy \cite{ATLAS:2016sno}.}.

Interestingly,  another hint of new physics appears in the 
ATLAS supersymmetric gluino searches at both 8 and 13~TeV where an on-shell $Z$-boson is produced together with jets and 
missing $E_T$ in a cascade decay \cite{ATLAS8dilepton,ATLAS13dilepton}. Motivated by this possibility, we address a question 
that has been investigated in the literature as to whether or not the production of VLQ states can simultaneously 
explain both this apparent signal as well as mediate the potential diphoton excess ({\it e.g.}, \cite{onZ-VLQ}). To address 
this issue in our study we perform a combined fit to both the ATLAS and CMS on-shell $Z$-boson + MET searches together with a large 
set of other supersymmetric particle search channels. We find that since the CMS searches for the same
final state \cite{CMS8dilepton,CMS13dilepton} do not see an equivalent signal, and that once these are combined no 
significant excess remains in the data.

Our study begins in Sec.~\ref{sec:model_details} where we present the details of our general model framework and report on 
the decay rates of the new particles which are relevant for LHC phenomenology. We go further in Sec.~\ref{sec:fit} to show 
the results of a dedicated fit of our model parameters accounting for the benchmark diphoton signal as well as the constraints arising 
from a large number of other LHC searches. In this section, we also investigate the possibility of explaining the ATLAS 
on-shell Z + MET excess by the pair production of VLQs. We summarize the results of our analyses in Section \ref{sec:summary} and in 
the Appendix~\ref{sec:appendix}, we provide the formulae for the components of the tree-levels decays of the singlet scalar 
that are induced by mixing with the SM Higgs. 

\section{Model details}
\label{sec:model_details}

The specific scenario that we consider is the Standard Model (SM) augmented by a real singlet scalar field together with 
one additional $SU(2)_L$ doublet of vector-like quarks (VLQs) which are allowed to mix with either the corresponding first 
or third generation SM quarks through the couplings with the usual Higgs doublet. 

\subsection*{The scalar potential}

The renormalizable scalar potential for the SM Higgs doublet $H$ plus an additional real singlet
scalar $S$ reads~\cite{Krasnikov:1992zk,O'Connell:2006wi,Barger:2007im,Barger:2008jx} 
\begin{align}
  V(H,S) = -\mu^2 H^\dagger H + \lambda (H^\dagger H)^2 - \frac{\mu_S^2}{2} S^2 + \frac{\lambda_S}{4} S^4 + \frac{\lambda_{SH}}{2} H^\dagger H S^2\,,\label{eq:scalar_potential}
\end{align}
with the real quartic couplings $\lambda$, $\lambda_S$ and $\lambda_{SH}$ and the mass terms $\mu^2$ and $\mu_S^2$. Both scalars can develop a vacuum expectation value (vev) and H and S can be expanded around their vacuum states as 
\begin{align}
  H = \bpm i \phi^+ \\ \frac{1}{\sqrt{2}} (h + v_H + i \phi^0) \epm \quad \text{and} \quad S = (s+v_S)\,,
\end{align}
where $v_H$ and $v_S$ are the corresponding vevs with 

\begin{align}
v_H^2=\frac{2\lambda_S\mu^2-4\lambda_{SH}\mu_S^2}{\lambda\lambda_S-4\lambda_{SH}^2},\\
v_S^2=\frac{2\lambda\mu_S^2-4\lambda_{SH}\mu^2}{\lambda\lambda_S-4\lambda_{SH}^2}\,.
\end{align}

Note that the scalar potential is $Z_2$ symmetric, but also that the Yukawa coupling of $S$ to the VLQs, 
$S \overline{\text{V}} \text{V}$ given below,
will explicitly break this symmetry. The radiatively generated 
terms $S$, $S^3$ and $S H^\dagger H$ all have loop suppressed couplings and can safely be neglected~\cite{Z2symmetry}. 
Using the minimization conditions of the scalar potential, the eigenvalues of the mass matrix of the scalar sector can be 
expressed in terms of the vevs and the Lagrangian parameters $\lambda$, $\lambda_S$ and $\lambda_{SH}$ as 

\begin{align}
M_{h_1,h_2}^2=\lambda v_H^2+\lambda_S v_S^2\mp\sqrt{(\lambda_S v_S^2-\lambda v_H^2)^2+\lambda_{SH}^2 v_S^2 v_H^2}\,.
\end{align}

The mixing of the interaction eigenstates $h$ and $s$ into the mass eigenstates $h_1$ and $h_2$ is described by
\begin{align}
 \bpm h_1 \\ h_2 \epm &= \bpm \cos\phi & -\sin\phi \\ \sin\phi & \cos\phi \epm \bpm h \\ s \epm\,,
\end{align}
with the mixing angle given by 
\begin{align}
 \tan(2\phi) = \frac{\lambda_{SH} v_S v_H}{\lambda_S v_S^2 - \lambda_H v_H^2}\,.
\end{align}

We define $\phi$ such that for $\phi=0$, the lighter eigenstate corresponds to the SM-like Higgs boson. The couplings of 
$h_1(h_2)$ to the SM particles are suppressed by $\cos\phi(\sin\phi)$ for non vanishing $\phi$ and thus the SM Higgs boson 
has reduced couplings. However, the Higgs coupling measurements are consistent with the SM expectation and hence a strict 
limit on the mixing angle can be derived from the LHC Higgs data. There are also direct limits from searches of heavy Higgs 
bosons at the LHC, in particular decays into the SM gauge bosons. In addition, the presence of a heavy scalar can influence 
electroweak precision data. In a closely related model, Ref.~\cite{simp_higgs} derives an upper limit on this mixing  
of $|\phi| \lesssim 0.35$.\footnote{Later we will see that we have to choose a very small value for this mixing angle 
which is well below these experimental limits in order to obtain a sufficiently large branching ratio into diphotons for the 
heavy scalar.}

In the following, we choose to work with the physical masses and mixing angles instead of the parameters of the scalar 
potential in Eq.~\eqref{eq:scalar_potential}. To be concrete, we choose the input parameters as 

\begin{align}
M_{h_1}=125 ~\text{GeV},\,M_{h_2}=750\,~\text{GeV},\,v_H=246\,\text{GeV},\,v_S,\,\phi\,.
\end{align}

The other parameters are then fixed by the mass formula and the minimization conditions. To conclude this subsection we
briefly discuss the perturbativity constraints on our quartic couplings. We perform our numerical analysis in terms 
of the phenomenological parameters such as the physical masses and the mixing angles. However, for specific values of the 
masses and the mixing angle, the parameters of the scalar sector can become non-perturbative. We have explicitly checked that 
all three quartic couplings given above are well below $\sqrt{4\pi}$ in our numerical analyses.

\subsection*{The fermion sector}

Here we consider only VLQs in $SU(2)_L$ doublet representations, further demanding that the VLQs can have Yukawa couplings to 
the corresponding SM fermions via the SM Higgs which then results in only three possible $SU(2)_L$ doublet representations 
~\cite{Dawson_VLQdoublets}; these are summarized in Table~\ref{tab:VLQdoublets}. After electroweak symmetry breaking (EWSB), 
the electric charges of the new VLQs that we consider are given by $Q_X = \frac{5}{3}$, $Q_T = \frac{2}{3}$, $Q_B = -\frac{1}{3}$ and 
$Q_Y = -\frac{4}{3}$. The weak and mass eigenstates associated with the electric charge $\frac{5}{3}$ and $-\frac{4}{3}$ states are equivalent 
since we assume that only one VLQ $SU(2)_L$ doublet is present in our scenario and there are no corresponding SM states with which
theses charges can mix. However, the new states $T$ and $B$ will mix with the up-type and down-type 
SM quarks, respectively. Here we will assume that this mixing involves {\it either} the first or the third generation quarks 
only. 

\begin{table}[htb]
\begin{tabular}{cc}
VLQ model & Representation \\ \toprule
$(X,T)$ & $(3,\,2,\,\frac{7}{6})$\\[2mm]
$(T,B)$ & $(3,\,2,\,\frac{1}{6})$\\[2mm]
$(B,Y)$ & $(3,\,2,\,-\frac{5}{6})$
\end{tabular}
\caption{Possible vector-like quark doublets assuming possible Yukawa couplings to SM quarks via the usual Higgs doublet 
and their hypercharges.}
\label{tab:VLQdoublets}
\end{table}

As a specific example we consider in detail the case of the $(X,T)$ doublet here. The other VLQ doublet cases follow in 
a straightforward manner. The most general gauge-invariant Yukawa and mass Lagrangian for the vector-like fermions and the 
SM up-type quarks of either the first or third generations is given by
\begin{align}
  \lag_\text{Yuk+mass} = \lambda_u \bar{Q}_L H u'_R + \lambda' \left(\bar{X}, \bar{T} \right)_L H u'_R 
  	+ \lambda'' \left(\bar{X}, \bar{T} \right)_L \bpm X \\ T \epm_R S 
  	+ m_D \left(\bar{X}, \bar{T} \right)_L \bpm X \\ T \epm_R + \text{h.c.}\,,\label{eq:Yukawa}
\end{align}
where the prime denotes the weak interaction eigenstates. The Lagrangian for the corresponding $(B,Y)$ model can be obtained 
upon the following substitutions: $X \rightarrow Y$, $T \rightarrow B$, $u \rightarrow d$ and $H \rightarrow \tilde{H}$, 
where $\tilde{H} = i \sigma_2 H^*$. For the $(T,B)$ doublet we need to substitute $T \rightarrow B$, $X\rightarrow T$ and 
$u \rightarrow d$ and also add two additional terms to the Lagrangian
\begin{align}
  \lag_\text{Yuk+mass} \supset  \lambda'_u \left(\bar{T}, \bar{B} \right)_L \tilde{H} u'_R +  m'_D \left(\bar{u}', \bar{d}' \right)_L \bpm T \\ B \epm_R  + \text{h.c.}\,.
\end{align}

For simplicity we consider the mass of the VLQs to be generated by Yukawa couplings to $S$ only, thus we set their both 
Dirac mass terms to zero $m_D = m'_D = 0$. The mass of the vector-like state $X$ can easily be read off as 
$m_X = m = \lambda'' v_S$. However, the vector-like $T$ and the SM up quark mix and we can write their mass term as
\begin{align}
  \lag_\text{Yuk+mass} \supset  \bpm \bar{u}, \bar{T} \epm_L \bpm \frac{\lambda_u v_H}{\sqrt{2}} & 0 \\[2mm]
  				\frac{\lambda' v_H}{\sqrt{2}} & \lambda'' v_S \epm 
  				\bpm u' \\ T' \epm_R 
  				=  \bpm \bar{u}, \bar{T} \epm_L \mat \bpm u' \\ T' \epm_R \,.
\end{align}
Here, in the case of mixing with the first generation $u$-quark, which we consider first, $\lambda_u$ is small compared to 
both $\lambda'$ and $\lambda''$ and can be neglected.

The diagonalization of the mass matrix in this case occurs via a biunitary transformation 
$\mat_\text{diag}= U_L^\dagger \mat U_R$ and leads to 
\begin{align}
  (U_L^\dagger \mat U_R)(U_R^\dagger \mat^\dagger U_L)  =&  U_L^\dagger \, \text{diag}(0, \lambda'' v_S + \frac{1}{2} \lambda'^2 v_H^2)  \, U_L  \label{eq:massMatrixVLQ_1},\\
  (U_R^\dagger \mat^\dagger U_L)(U_L^\dagger \mat U_R)  =&  U_R^\dagger \mat^\dagger \mat U_R = U_R^\dagger  
  		\bpm \frac{\lambda'^2 v_H^2}{2} & \frac{\lambda' v_H m}{\sqrt{2}} \label{eq:massMatrixVLQ} \\[2mm]
  		\frac{\lambda' v_H m}{\sqrt{2}} & m^2 \epm U_R\,.  		
\end{align}
Since the mass matrix in Eq.~\eqref{eq:massMatrixVLQ_1} is already diagonal, we have $U_L = \mathbb{1}$, and only the 
right-handed particles mix. The corresponding mixing angle, $\theta_R$, is given by
\begin{align}
  \tan(2\theta_R) = -2 \frac{\frac{1}{\sqrt{2}} \lambda' v_H m}{m^2-\frac{1}{2}\lambda' v_H^2}\,.
\end{align}
For VLQs mixing with instead the third generation of SM quarks, $\lambda_u$ is no longer much smaller than $\lambda'$ and 
$\lambda''$, however the right-handed mixing angle still dominates over the left-handed one, particularly for mixing in 
the bottom sector~\cite{Aguilar-Saavedra:2013qpa}
\begin{align}
  \tan(\theta_L) = \frac{m_q}{m_\text{VLQ}} \tan(\theta_R)\,.
\end{align}
Due to the CKM constraints~\cite{Alok_mixing_angle}, the mixing with the first generation 
SM quarks has to be $\ord(0.01)$ or less if mixing between the VLQ and left-handed 
SM fields is allowed. The mixing with the right handed first generation SM quarks is 
bounded by the couplings of the $Z$-boson to light quarks which has precisely been measured 
by LEP \cite{ALEPH:2005ab} and the precision measurements of the atomic parity 
violation experiments \cite{Deandrea:1997wk}. Using these constraints, limits can be computed
on the mixing with right-handed first generation quarks and this must be 
smaller than $\mathcal{O}(0.1)$ \cite{Buchkremer:2013bha}.

Limits can also be derived on the third generation mixing angles 
since the presence of VLQ quarks modifies the oblique parameters $S$ and $T$ 
\cite{Peskin:1990zt} as well as the $Zb\bar b$ coupling \cite{Bamert:1996px}. 
In the case of the model with a dominant right handed mixing angle for $(T,B)$,
the mixing angle cannot be larger than $\mathcal{O}(0.1)$ \cite{Aguilar-Saavedra:2013qpa}
and similar constraints are also present for the $(X,T)$ and $(B,Y)$ doublet scenarios. 

We note that assuming masses for VLQ states of the order of 800 GeV (as motivated by direct 
LHC searches), the left-handed mixing angle is already subdominant compared to the
right-handed mixing angle even for the top sector. In fact the only effect of the left-handed
mixing angle on the phenomenology explored in this study is to modify the branching ratios by 
$\mathcal{O}$(few \%). For VLQ mixing with the SM top quark, the left-handed mixing angle  
of SM quarks is included in the numerical results.
In our analysis we will keep the mixing angle fixed at $\theta_R = 0.01$ for the mixing 
with either the first or third generation, since its variation below this value will 
not notable affect the phenomenology of our model.\\

%

Upon diagonalization the mass of the heavy new quark $T$ is given by $m_T =
\sqrt{m^2 + \frac{1}{2}\lambda'^2 v_H^2} \approx m_X (1 + \frac{1}{2}\theta_R^2)$ so that it would appear that $T$ is heavier 
than $X$ by $\sim 50$ MeV if $m_X \sim 1 $ TeV. At the 1-loop level, however, the masses of the vector-like quarks receive 
electroweak radiative corrections, leading instead to $m_X > m_T$, but with $\Delta m = m_X - m_T < 1 ~\gev$ (see Eq.~(12) 
of~\cite{Cirelli:2009uv}):
\begin{align}
  m_X-m_T &= \frac{\alpha s_W^2 m}{4\pi^2} \, \frac{5}{3} \, h\left(\frac{M_Z}{m}\right) 
  	\approx \frac{5\alpha s_W^2 }{6\pi}  M_Z \approx 0.57 \, \gev,
            \quad {\rm with}\quad h(r)\approx 2 \pi r\,,
\end{align}
where $s_W = \sin\theta_W$ is the sine of the SM weak mixing angle.

Let us now consider the decays of $X$ and $T$. Because of its charge, $X$ can only decay via a $W$ boson. The possible
decay modes are $X \rightarrow T W^*$ and $X \rightarrow u_R W$. Although the latter mode is suppressed by mixing, it 
strongly dominates over the decay to a $T$, because of the small mass difference of the vector-like quarks given above 
\begin{align}
  \Gamma(X \rightarrow T l \nu) &= \frac{G_F^2 (\Delta m)^5}{15 \pi^3} < 10^{-12}\, \gev \text{ at $c_R=$1 }\,, \\
  \Gamma(X \rightarrow u_R W)   &= \frac{G_F m_X^3}{8 \sqrt{2}\pi} s_R^2 (1-r_W)^2 (1+2r_W) \approx 1.13\cdot 10^{-4} 
  	\left(\frac{m_X}{700~GeV} \right)^3 \left(\frac{s_R}{10^{-3}} \right)^2 \gev\,,
\end{align}
where $r_i = m_i^2/m_X^2$ and we denote the sine and cosine of the mixing angle $\theta_R$ by $s_R = \sin\theta_R$ 
and $c_R =\cos\theta_R$, respectively.{\footnote {A corresponding, but somewhat more complex expression exists when $X$ 
decays instead to $t_RW$.}} The vector-like $T$ can decay to either the SM Higgs or to a $Z$ boson plus a SM up-type quark. 
As long as the mass of the vector-like quarks are well below the mass of the new scalar singlet, $T$ will decay to 
$T\rightarrow uZ$ and $T \rightarrow uh_1$ with a branching fraction of approximately $50$\%
each, irrespectively of the family to which the VLQs couple. In the case of decays to first generation quarks we find 
\begin{align}
  \Gamma(T \rightarrow uZ)   &= \frac{G_F m_T^3}{16 \sqrt{2}\pi} s_R^2 c_R^2 (1-r_Z)^2 (1+2r_Z)\,,\\
  \Gamma(T \rightarrow u h_1)&= \frac{G_F m_T^3}{16 \sqrt{2}\pi} s_R^2 c_R^2 \left(c_\phi + \frac{v_H}{v_S} s_\phi \right)^2 
  	(1-r_{h_1}) ^2\,.
\end{align}
Because of the assumed absence of a left-handed mixing angle, there is no decay of the $T$ into a $W$ boson. 
When $m_T > M_{h_2}$, the decay mode $T\rightarrow h_2 u$ opens with a decay width of (again assuming decay to the first 
generation) 
\begin{align}
  \Gamma(T \rightarrow u h_2)&= \frac{G_F m_T^3}{16 \sqrt{2}\pi} s_R^2 c_R^2 \left(s_\phi - \frac{v_H}{v_S} c_\phi \right)^2 
  	(1-r_{h_2}) ^2\,.
\end{align}

We conclude this subsection with a discussion of the $(T,B)$ model which has four mixing angles in the most 
general case. Again, the left-handed mixing angles are suppressed especially for the case of mixing with 
the first generation and hence the left-handed mixing angles 
are not considered as mentioned before. The branching ratios of $T$ and $B$ depend on 
the mixing angles $\theta_R^\text{up}$ and $\theta_R^\text{down}$, {\it e.g.}, the $T$ 
quark does not couple to the $Z$ and $H$ boson when $\sin\theta_R^\text{up}$=0. Here, we  
consider a scenario with one mixing angle for the sake of simplicity, and thus we 
set $\theta_R^\text{up} = \theta_R^\text{down} = \theta_R$. With this assumption,
the couplings of both VLQs in the $(T,B)$ doublet representation to the $Z$ are of the 
same strength and proportional to $\sin\theta_R\cos\theta_R$. The same conclusion can 
be drawn for the couplings to the $W$ boson which are also proportional 
to $\sin\theta_R\cos\theta_R$  \cite{Aguilar-Saavedra:2013qpa}. For the Higgs couplings, the relevant
interaction terms of the Lagrangian is given by \cite{Aguilar-Saavedra:2013qpa}
\begin{align}
\bar q\left(Y_{qQ}^L P_L+Y_{qQ}^R P_R\right)Q H +h.c.\,.
\end{align}
Here, the couplings $Y^L_{tT}$ and $Y^L_{bB}$ are of the same size, but $Y^R_{qQ}$ differs 
by a factor $\frac{m_q}{m_{Q}}$ for the top and bottom quark, respectively.


\subsection*{Loop-induced Effective couplings}

The heavy scalar $h_2$ is mainly produced in gluon-fusion with the vector-like quarks running in the fermion loop assuming 
a small mixing angle $\phi$. At vanishing mixing with the SM Higgs, all the decay modes of $h_2$ (which then coincides with 
$s$) are purely loop-induced assuming the on-shell decays to the VLQs are closed. The interaction of the new scalar with 
the vector bosons is given by the effective Lagrangian
\begin{align}
  \lag_\text{eff} &= - \frac{1}{4} g_{sgg} s G_{\mu \nu} G^{\mu \nu} - \frac{1}{4} g_{s\gamma\gamma} s A_{\mu \nu} A^{\mu \nu}
  	- \frac{1}{4} g_{sZZ} s Z_{\mu \nu} Z^{\mu \nu} - \frac{1}{2} g_{sWW} s W_{\mu \nu} W^{\mu \nu}
  	- \frac{1}{2} g_{sZ\gamma} s A_{\mu \nu} Z^{\mu \nu}\,,
\end{align}
where $g_{sxy}$ denotes the effective coupling to the vector bosons $x,y$. For gluons and photons these are given by 
\begin{align}
  g_{sgg} &= \frac{\alpha_s}{4\pi}  \sum_{i=X,T} \frac{g_{sii}}{m_i} \, A_{1/2}(\tau_i)\,, \label{eq:beginBRcalc} \\
  g_{s\gamma\gamma} &= \frac{\alpha}{4\pi}  \left( 2 N_C \sum_{i=X,T} Q_i^2 \frac{g_{sii}}{mi}\, A_{1/2}(\tau_i)  \right)\,,
\end{align}
where $A_{1/2}$ is the standard loop integral
\begin{align}
	A_{1/2}(\tau) &= 2 \tau \; [1+(1-\tau)\; f(\tau)]\,, \\
	f(x) &= \arcsin^2(1/\sqrt{x})\,,
\end{align}
with $\tau_i = 4 m_i^2 / m_{s}^2$. The couplings of the scalar $s$ to the VLQs are given by 
\begin{align}
	g_{sTT} &= \frac{m_T}{v_H} \left(s_R^2 s_\phi + \frac{v_H}{v_S} c_R^2 c_\phi\right)\,,\\
	g_{sXX} &= \frac{c_\phi c_R m_T}{v_S}\,.
\end{align}
Neglecting the squares of the mass of the $W$ and the $Z$ bosons relative to that of the $s$, the effective couplings of 
the scalar $s$ to $Z\gamma$, $ZZ$ and $WW$ are given by \cite{Benbrik:2015fyz,Rays_of_light}
\begin{align}
  g_{sZZ} &= \frac{\alpha}{4\pi}  \left[ \frac{2 N_C}{s_W^2 c_W^2} 
  			\sum_{i=X,T} (I^3_i - s_W^2 Q_i)^2 \frac{g_{sii}}{m_i}\, A_{1/2}(\tau_i)  \right]\,,\\
  g_{sZ\gamma} &= \frac{\alpha}{4\pi}  \left[ \frac{2 N_C}{s_W c_W} 
  			\sum_{i=X,T} Q_i (I^3_i - s_W^2 Q_i) \frac{g_{sii}}{m_i}\, A_{1/2}(\tau_i)  \right]\,,\\
  g_{sWW} &= \frac{\alpha}{4\pi}  \left[ \frac{N_C}{2s_W^2} 
  			\sum_{i=X,T} \frac{g_{sii}}{m_i}\, A_{1/2}(\tau_i) \right]\,. \label{eq:endBRcalc}
\end{align}

In Table \ref{tab:BRs} we give some values for the resulting branching ratios of $s$ for the different VLQ doublets.
We take into account a K-factor of $1.638$ \cite{Liu:2016mpd} for the production of $s$ via gluon fusion at $13 \,\tev$ and a 
K-factor of $1 + 67\alpha_s(M_{h_2})/4\pi$ for the decay into two gluons.{\footnote {We employ the results 
of Ref. \cite{Djouadi:1991tka} with $\mu=m_s$ and $N_f=6$ which is applicable when $m_t^2/m_s^2<<1$ and 
$4m_{VLQ}^2/m_s^2>1$.}}

Additional decay modes can be induced by the mixing with the SM Higgs boson and these lead to a  
tree level contribution to $WW$ and $ZZ$ final states. We provide the explicit formulae 
for the component of the tree-level decays induced by mixing with the SM Higgs in the 
Appendix \ref{sec:appendix}. We also note that while our formulae neglect the interference
between the loop-induced couplings and those arising from mixing with the SM Higgs, all of our 
numerical results include such effects.
An additional contribution could also come from the mixing with the SM-quarks, but in 
our case this effect is negligible due to the small values of the relevant mixing angles $\theta_{L,R}$.

For the numerical analysis, we implemented the three models using \toolfont{FeynRules2.3.18}~\cite{FeynRules,Christensen:2008py}. 
We validated our implementation by comparing our results to the kinematic distributions as depicted in
the ATLAS search for VLQs \citep{ATLAS13VLQ}.


\begin{table}[htb]
\begin{tabular}{ccccccccc}
VLQ model & Representation & $\gamma Z/\gamma\gamma$ & $ZZ/\gamma\gamma$ & $WW/\gamma\gamma$
& $gg/\gamma\gamma$ & $\Gamma_{s \rightarrow \gamma\gamma}$ [MeV] & $\Gamma_\text{Tot}$ [MeV] & $R_{\gamma\gamma}$ [fb]\\ \toprule
$(X,T)$ & $(3,\,2,\,\frac{7}{6})$ & $0.07$ & $0.59$ & $0.90$ & $17.0$ & $1.03$ & $20.0$ & $6.2$\\[2mm]
$(T,B)$ & $(3,\,2,\,\frac{1}{6})$ & $5.02$ & $9.11$ & $30.2$ & $570.3$ & $0.03$ & $18.8$ & $0.2$\\[2mm]
$(B,Y)$ & $(3,\,2,\,-\frac{5}{6})$ & $0.01$ & $1.21$ & $2.61$ & $49.3$ & $0.35$ & $19.1$ & $2.2$
\end{tabular}
\caption{Branching ratios into various final states of the new scalar $s$, the total 
width $\Gamma_\text{Tot}$ and the diphoton production rate $R_{\gamma\gamma}$ for 
the following input parameters: $v_S = 750$ GeV, $\phi = 0$, $m_\text{VLQ}=1$ TeV, 
$\sin^2\theta_w=0.2315$, $\alpha_s(M_s)=0.09036$.}
\label{tab:BRs}
\end{table}
\section{Results}
\label{sec:fit}
\begin{figure*}[ht!]
	\includegraphics[width=0.49\textwidth]{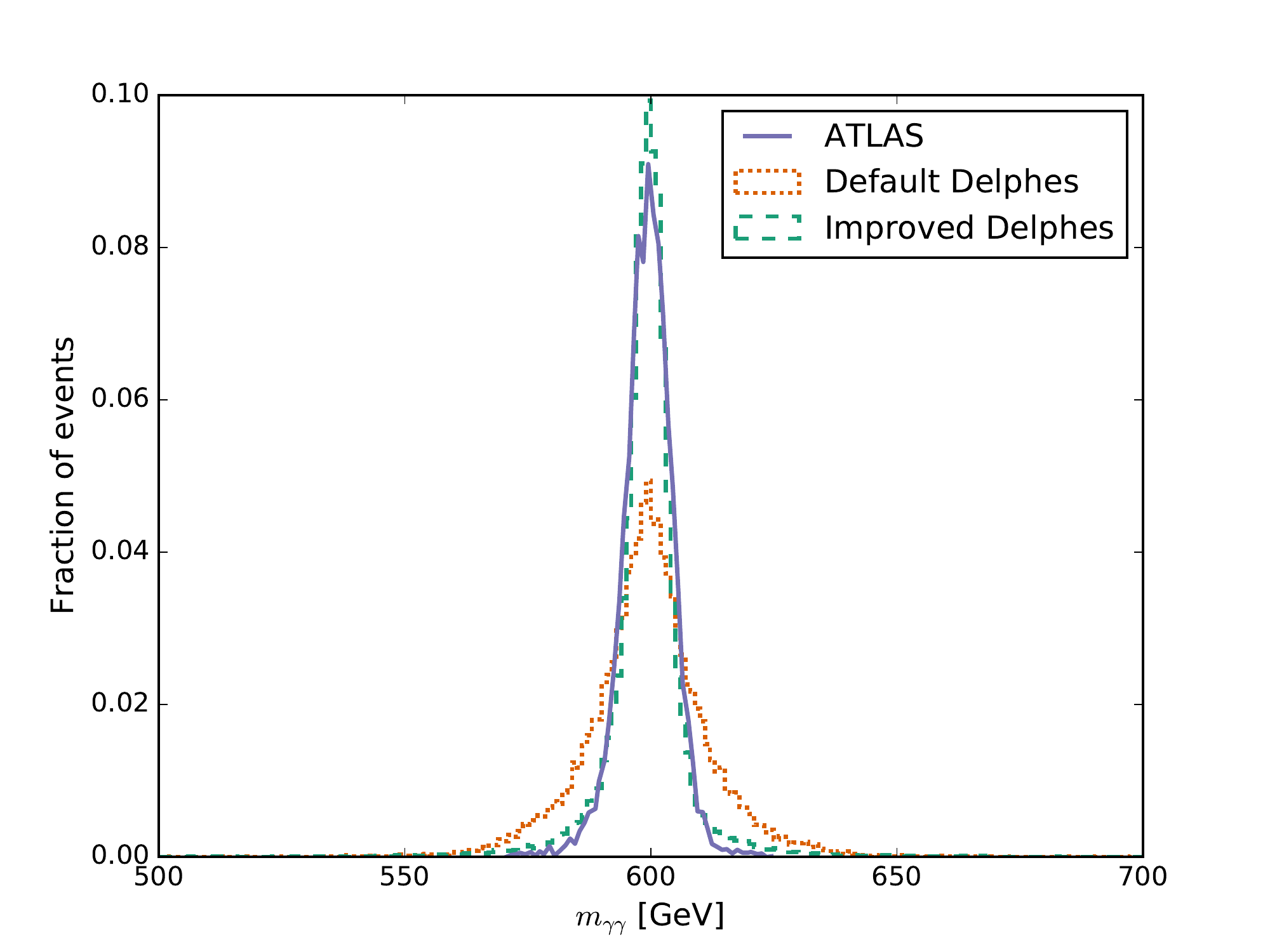}
	\;
	\includegraphics[width=0.49\textwidth]{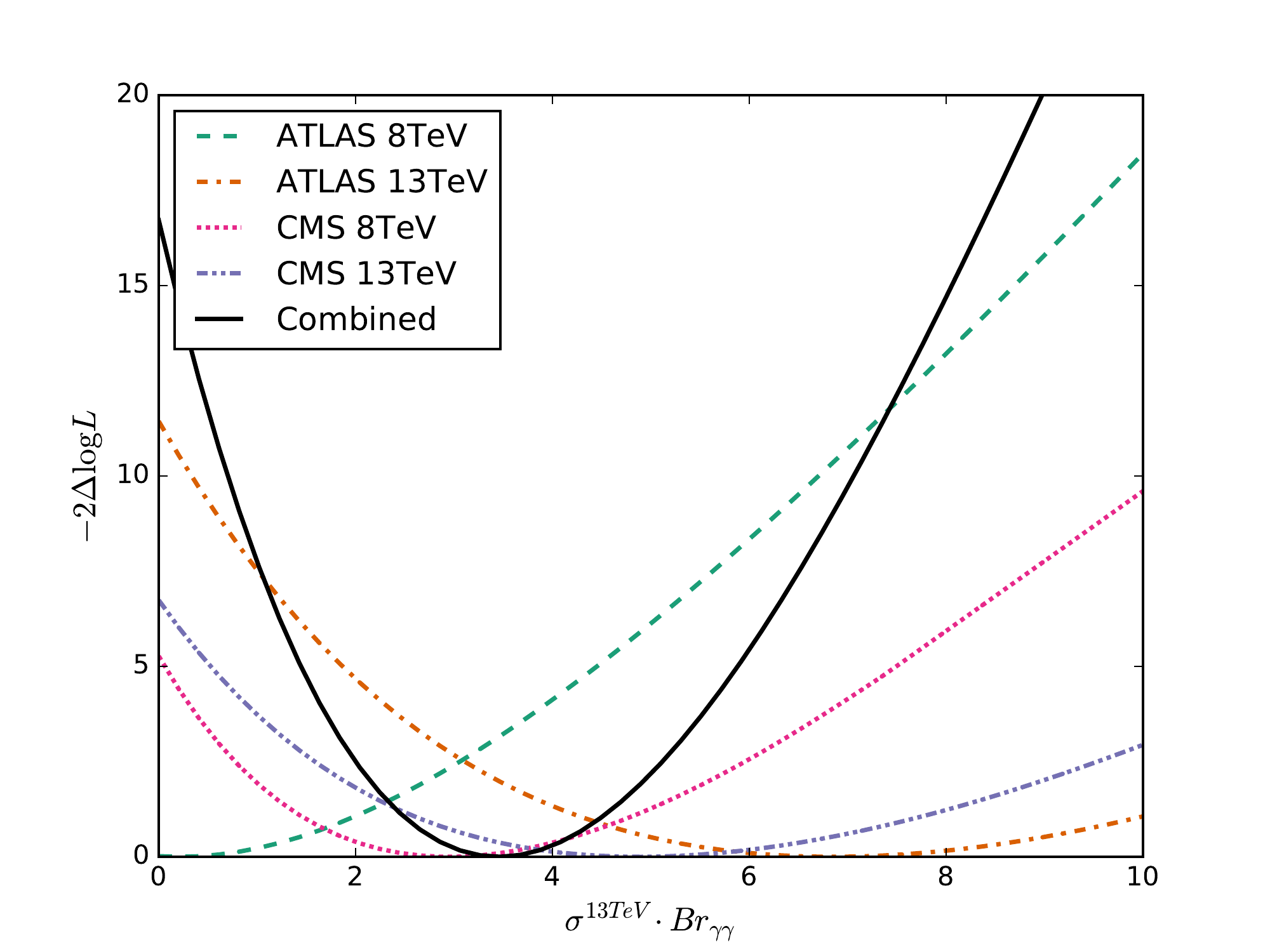}
	\caption{Detector resolution for the ATLAS detector compared with that from Delphes before and after tuning (left). 
	$\chi^2$ of the combined diphoton results as a function of the $\sigma\times\text{BR}(h_2 \rightarrow \gamma\gamma)$. The best 
	fit point lies at $\sigma\times\text{BR} = 3.5 \pm 1.0 \text{ fb}$. The CMS results are taken from Figure 10 (left) 
	in  \cite{Khachatryan:2016hje}.}
	\label{fig:widthATLAS}
\end{figure*}

We perform a combined fit for the diphoton signal rate and a number of ATLAS and CMS searches which are sensitive 
to VLQ pair production. For purposes of demonstration, we employ the observed value of the excess diphoton rate at 750 GeV.
To compare the results for the pair production of VLQs with a large number of LHC searches we make 
use of the tool \toolfont{CheckMATE} \cite{Drees:2013wra,Kim:2015wza} including the \toolfont{Delphes~3}
detector simulation \cite{deFavereau:2013fsa}. For all analyses, the anti-$k_T$ jet algorithm is used 
\cite{Cacciari:2011ma,Cacciari:2005hq,Cacciari:2008gp} and we also include additional, externally implemented analyses 
\cite{Cao:2015ara}. For each of the various signal regions we compute a likelihood by assuming that the various 
systematic uncertainties are distributed according to a Gaussian probability density function (PDF) and combine this with 
the Poisson distributed statistical uncertainty. Each analysis likelihood (including the diphoton results) are then combined 
to give a total $\chi^2$ for the model under test. When combining LHC analyses, we make sure to only include orthogonal 
signal regions.

To present our results we display the $\chi^2$ relative to that predicted by the SM alone,
\begin{align}
\chi^2_\text{rel} = -2 \left(\ln L - \ln L_\text{SM} \right)\,.
\end{align}

As discussed below, we also investigate all other final states that may be produced by a 750~GeV
resonance (as shown in Table~\ref{tab:limits}). However we find that in the models we explore here, none of these are relevant with 
the currently collected data since the current limits lie far from the anticipated rates.

\subsection{Fitting the diphoton signal rate}
%

\begin{table*}[ht!]
  \begin{tabular}{cc}
	Parameter & Range  \\ \toprule
	$m_\text{VLQ}$ [GeV] & $[600,1800]$ \\[1mm]
	$v_S$ [GeV] & $[750,2000]$\\[1mm] 
	$\phi$ & $[-0.1, 0.1]$ \\ 
	\bottomrule
  \end{tabular}
  \caption{Range of the parameters varied in the fit.}
  \label{tab:rangeParameters}
\end{table*}

\begin{figure*}[ht!]
	\includegraphics[width=0.49\textwidth]{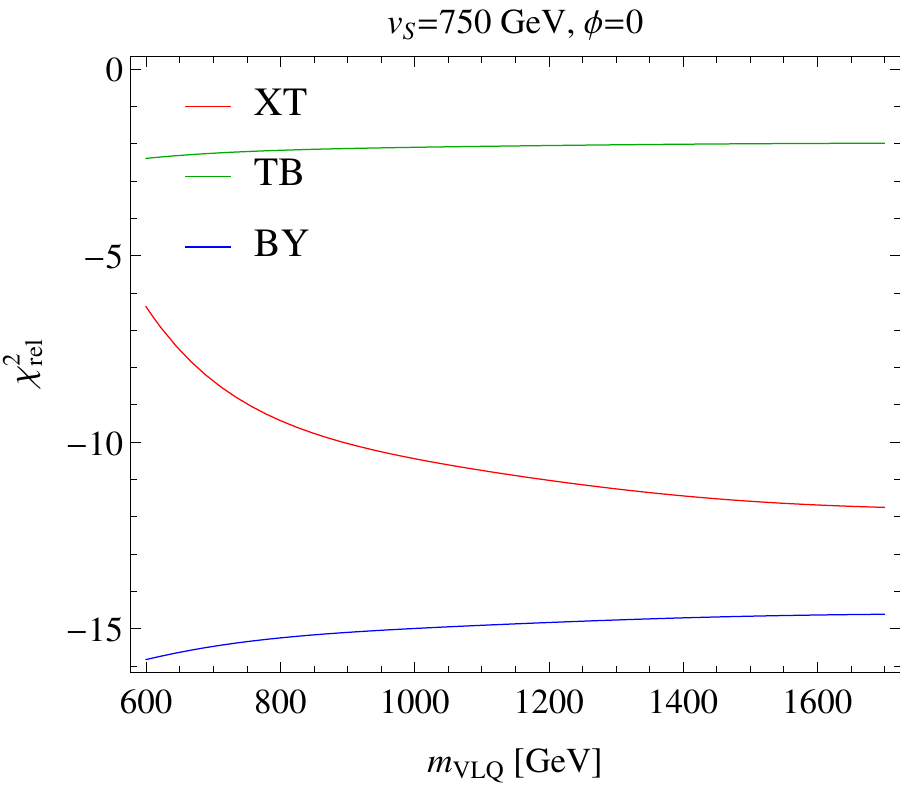}
	\caption{Comparison of the diphoton fit $\chi^2$ contributions for the different VLQ doublets.}
	\label{fig:diphotonLikeli}
\end{figure*}

\begin{figure*}[ht!]
	\includegraphics[width=0.49\textwidth]{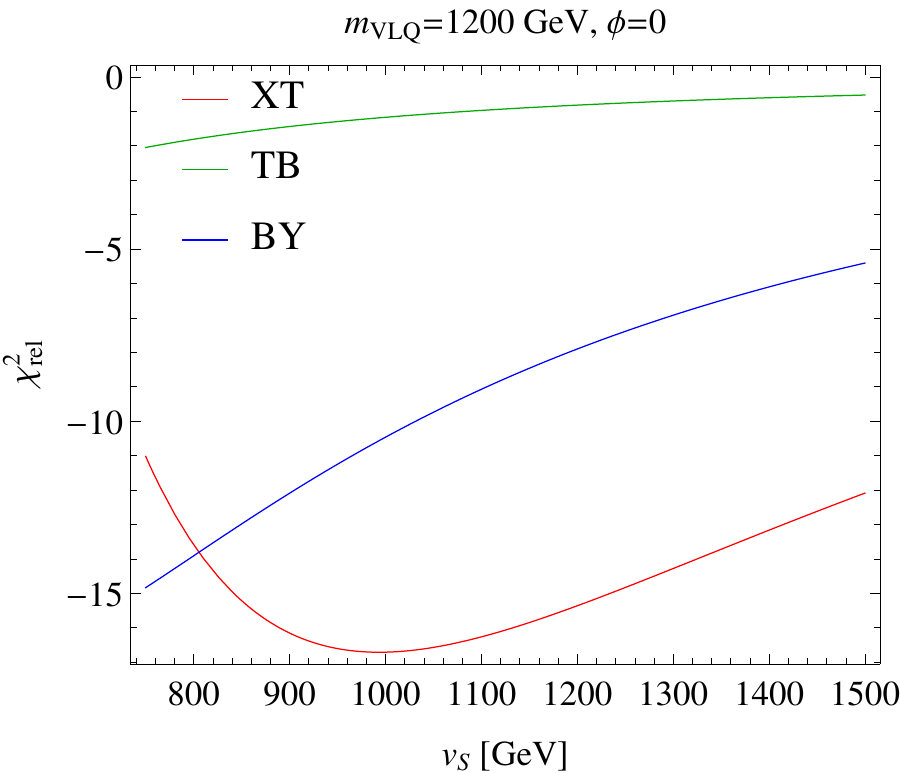}
	\;
	\includegraphics[width=0.49\textwidth]{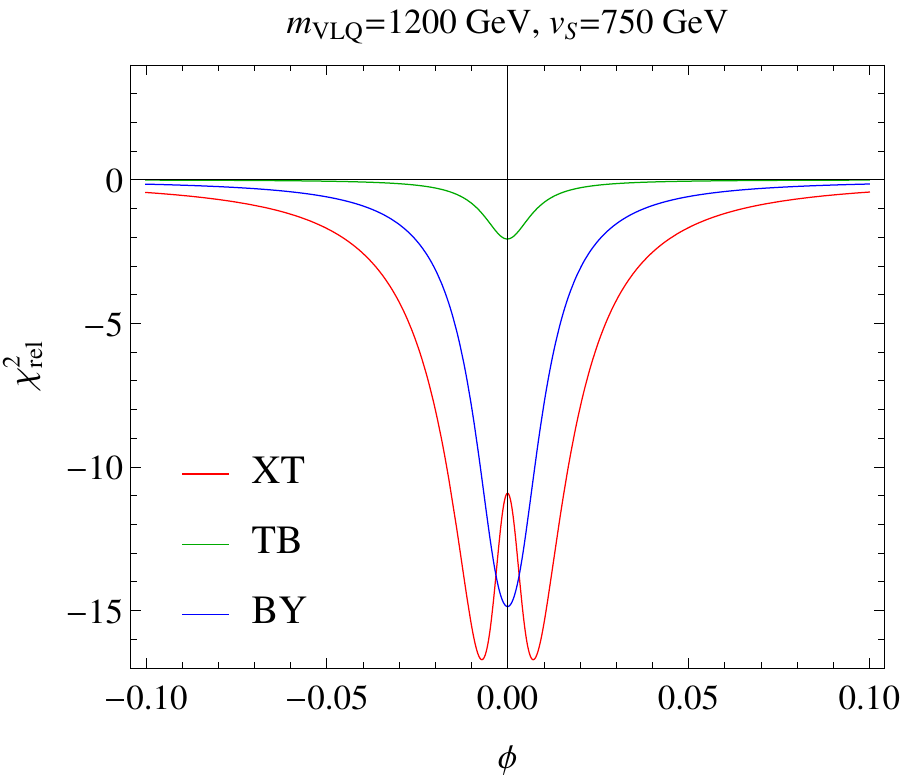}
	\caption{Comparison of the fits of vector-like quark doublets for the diphoton likelihood. Variation of the vev of the new scalar $v_S$ (left) and its mixing angle with the SM Higgs $\phi$ (right).}
	\label{fig:compare_vS_phi_first_gen}
\end{figure*}

We start our analysis of the different VLQ doublets by fitting the likelihood of the excess diphoton signal without taking into 
account other experimental constraints. We consider the diphoton searches by ATLAS and CMS at $13\,\tev$ and 
the $8 \,\tev$ results scaled to the $13\,\tev$ cross-section \cite{Aaboud:2016tru,Khachatryan:2016hje} in the invariant 
mass range $m_{\gamma\gamma}\subset [650,850] \,\gev$. 
We implemented the ATLAS analyses in \toolfont{CheckMATE} and tuned the Delphes detector card to 
better reproduce the detector response to photons which is over-smeared in the default Delphes setup, as seen in Fig.~\ref{fig:widthATLAS}. 
The likelihood for CMS is fitted to the values shown in \cite{Khachatryan:2016hje}. For a scalar 
resonance, we find the best fit point for the diphoton rate is $\sigma\times\text{BR} = 3.5 \pm 1.0 \text{ fb}$ 
(see Fig.~\ref{fig:widthATLAS}) and, due to the improved photon response that we have implemented, this value 
is somewhat below the one commonly shown in the literature \cite{Buckley}.\\

\begin{figure*}[t!]
	\includegraphics[width=\textwidth]{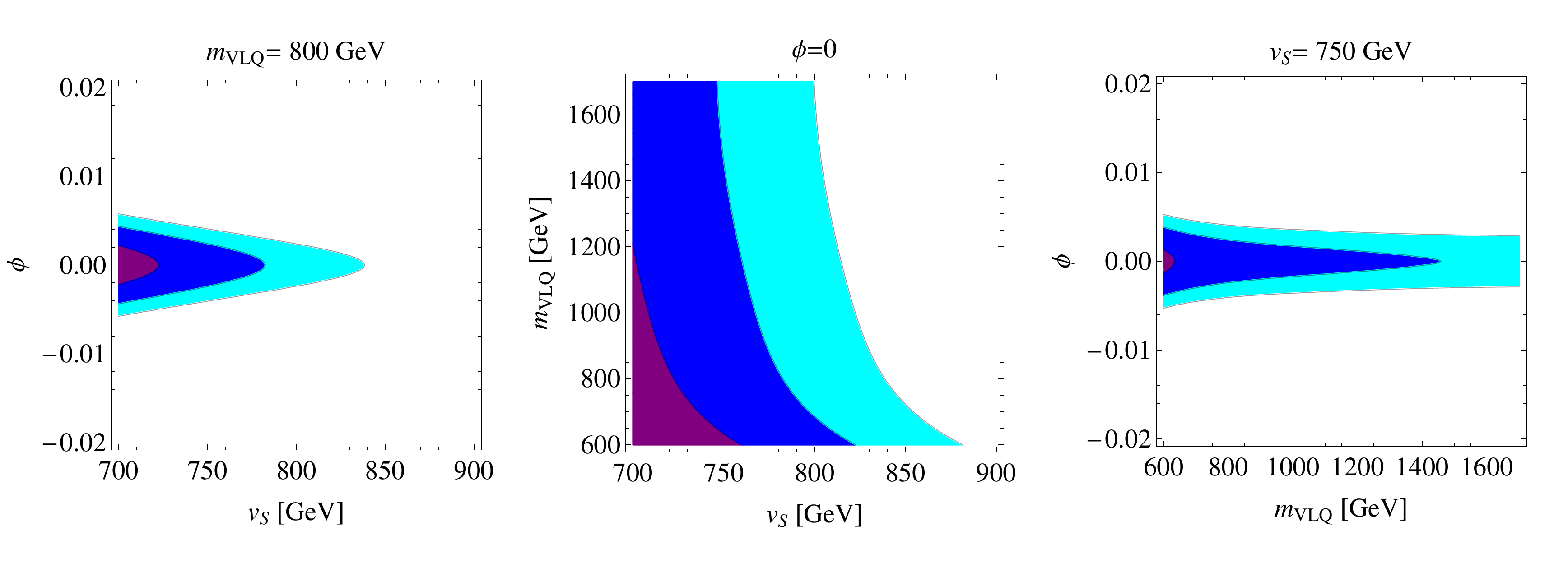}
	\includegraphics[width=\textwidth]{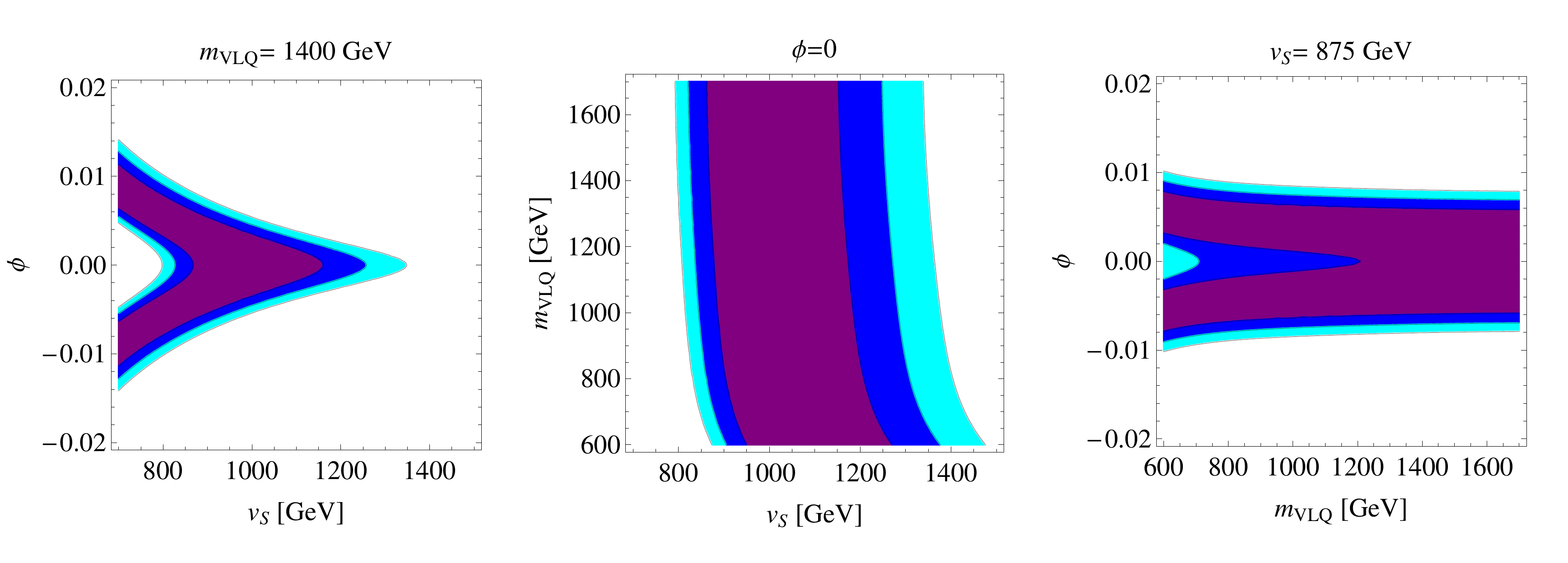}
	\caption{Contours of the likelihood for the $(B,Y)$ (upper) and $(X,T)$ (lower) model for the diphoton rate 
	contribution to the overall $\chi_{rel}^2$ only. The contours are shown for the $1\sigma$ (purple), $2\sigma$  (dark blue) 
	and $3\sigma$  (light blue) regions. Please note the different plot ranges and values for $v_S$ and $m_\text{VLQ}$.}
	\label{fig:contours}
\end{figure*}

The fit of the diphoton rate is independent of which family of SM quarks primarily couple to the VLQs since
we only consider production via gluon fusion. Keeping 
the mixing angle of the quark sector $\theta_R$ fixed (which as stated earlier, has to be small due to CKM constraints), 
the mass of the VLQs, $m_\text{VLQ}$, the vev of the new scalar $v_S$ and the scalar mixing angle $\phi$ are the 
parameters that influence the cross section times branching ratio into a pair of photons. For our fit, we vary those parameters 
in the ranges given in Table \ref{tab:rangeParameters}.

We calculate the production cross section of the new scalar $h_2$
using \toolfont{MadGraph} \cite{madgraph} including the K-factor of $1.638$ as mentioned above and calculate 
its BRs analytically via Eqs. \eqref{eq:beginBRcalc}-\eqref{eq:endBRcalc} and the equations 
given in App.~\ref{sec:appendix}. In Fig.~\ref{fig:diphotonLikeli} the dependence of $\chi^2_\text{rel}$ on the 
VLQ mass is given for the different VLQ doublets for $v_S=750$~GeV and $\phi=0$.
From the behavior of the curves one can see that the cross section times branching ratio for 
the $(T,B)$ and the $(B,Y)$ model is too small for the given parameter range, whereas in
contrast the $(X,T)$ model predicts a too large rate for the diphoton resonance.\\

As we allow $\phi$ to vary, the tree level decays $h_2 \rightarrow WW/ZZ/h_1h_1/f\bar{f}$ become possible 
(and relevant), suppressing the BR of the new scalar to two photons (see Appendix \ref{sec:appendix} for 
the tree-level decay widths of $h_2$). In addition, in order not to make the Yukawa couplings of the VLQs too large, 
we only consider the regime $v_S > 750\,\gev$. Since a larger $v_S$ corresponds to a smaller Yukawa coupling
(for fixed $m_\text{VLQ}$), increasing the value of $v_S$ will also lead to a smaller diphoton rate. 

Thus increasing either $v_S$ or $\phi$ in the given range will only worsen the quality of the diphoton 
fit for both the $(T,B)$ and $(B,Y)$ models, see Fig.~\ref{fig:compare_vS_phi_first_gen}. However, we note that 
for very low mass VLQs ($m_\text{VLQ}\sim 600$~GeV) the diphoton rate as observed by ATLAS and LHC can be 
reached within $1\sigma$ for the $(B,Y)$ model as shown in the upper panel of Fig. \ref{fig:contours}.

In contrast to the two models already discussed, when $v_S=750$~GeV and $\phi=0$, the $(X,T)$ model
actually predicts a too large diphoton rate. Consequently the fit can actually now
be improved with the variation 
of $v_S$ and $\phi$ (see lower panel of Fig.~\ref{fig:contours}). For example, for smaller values of $v_S$ or 
$m_\text{VLQ}$, a non-zero value of $\phi$ is found to be favored as a way to reduce the diphoton rate and
fit the experimental combined result. 


\subsection{LHC Constraints From Searches}
Additional constraints on the parameters of our models may arise from limits of the decay of the new scalar 
to final states other than $\gamma\gamma$ as well as from LHC searches that are sensitive to VLQ pair production. 
Since the mixing angle of the VLQs to SM quarks has to be very small, we find that there are no relevant 
constraints on our model parameters originating from the single production of VLQs.

We first consider the limits arising from the resonant searches in other relevant final states at 750~GeV. 
These are presented in Table \ref{tab:limits} and we find that all of these bounds are presently too weak 
to place constraints on any of the above models given the parameter ranges that we explore. \\

\begin{table*}[ht]
  \begin{tabular}{cc}
	Final state & $95 \%$ CL upper limit on $\sigma \times $ BR [fb] \\ \toprule
	$WW$ \cite{WWupperlimit}& $180$ \\
	$ZZ$ \cite{ZZupperlimit}& $55$ \\
	$jj$ \cite{jjupperlimit}& $2000$ \\
	$Z\gamma$ \cite{Zgamupperlimit}& $18$ \\
	$hh$ \cite{hhupperlimit} & $198$ \\ \bottomrule

  \end{tabular}
  \caption{$95 \%$ CL upper limits on $\sigma \times $ BR for different final states. None of the these are
  found to be relevant for the set of models and parameters investigated in this study.}
  \label{tab:limits}
\end{table*}


In contrast, the existing limits on the pair production of VLQs can set meaningful constraints
on the parameter space. We should first remember that the pair production of vector-like quarks is independent 
of the properties of the scalar sector $v_S$ and $\phi$ in the range relevant to fit the diphoton 
rate ($\phi \subset [-0.03,0.03]$, $v_S \subset [750,1500]$) since their branching ratios depend only 
very weakly on $v_S$ and $\phi$. As long as we keep the mixing angle of the quark sector $\theta_R$ fixed, 
the mass of the VLQs is therefore the only parameter influencing the results for VLQ pair production.

To compare the results for the pair production of VLQs with the set of LHC searches we consider, we again  
make use of \toolfont{CheckMATE}. In Table \ref{tab:listAnalysis} we display the list of the analyses that are 
sensitive to VLQ pair production and are taken into account as part of the present study. We include three $8 \, \tev$ 
and seven $13 \, \tev$ searches, which when combined corresponds to $105$ signal regions in total. For the 95\% CLs exclusion
we only consider the signal region with the best expected sensitivity and then apply the result using 
the collected data. When performing the combined fit of analyses, if an analysis has overlapping signal
regions we only include the signal region that was expected to be most sensitive from this analysis. Such
a procedure applies to all the ATLAS analyses apart from the VLQ search which has orthogonal signal regions. In addition, 
for analyses that target similar final states, each signal region was examined to make sure that no signal
regions overlap. For the final combination we assume no correlations in the systematic uncertainties between signal
regions which implies our exclusion is more conservative than if such correlations were included.

\begin{figure}[ht!]
	\includegraphics[width=0.49\textwidth]{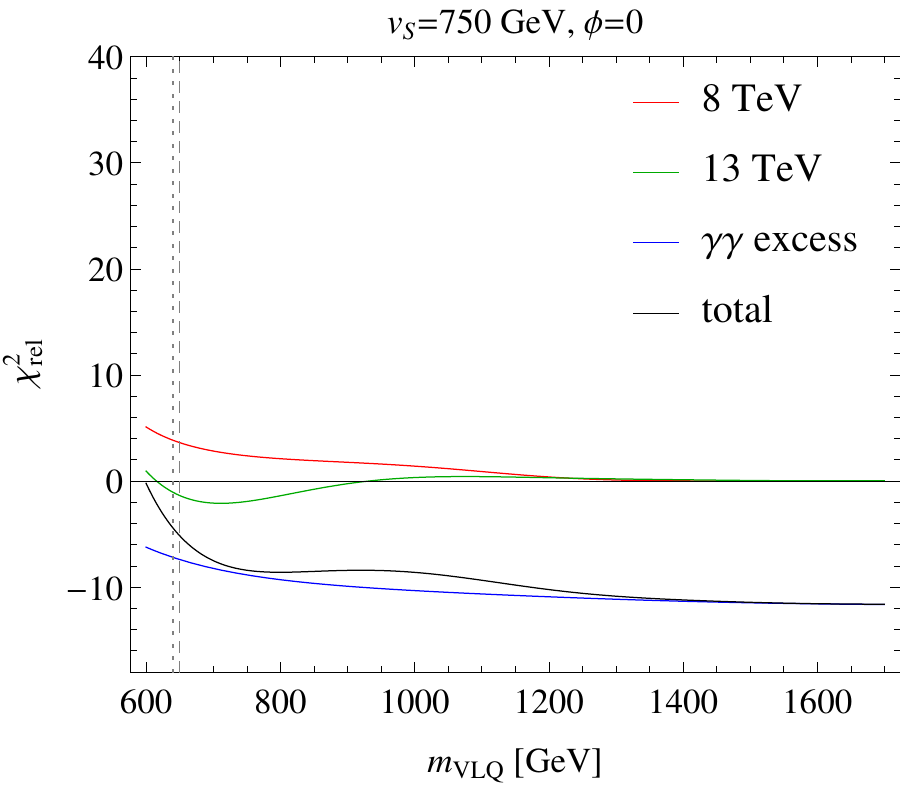}
	\;
	\includegraphics[width=0.49\textwidth]{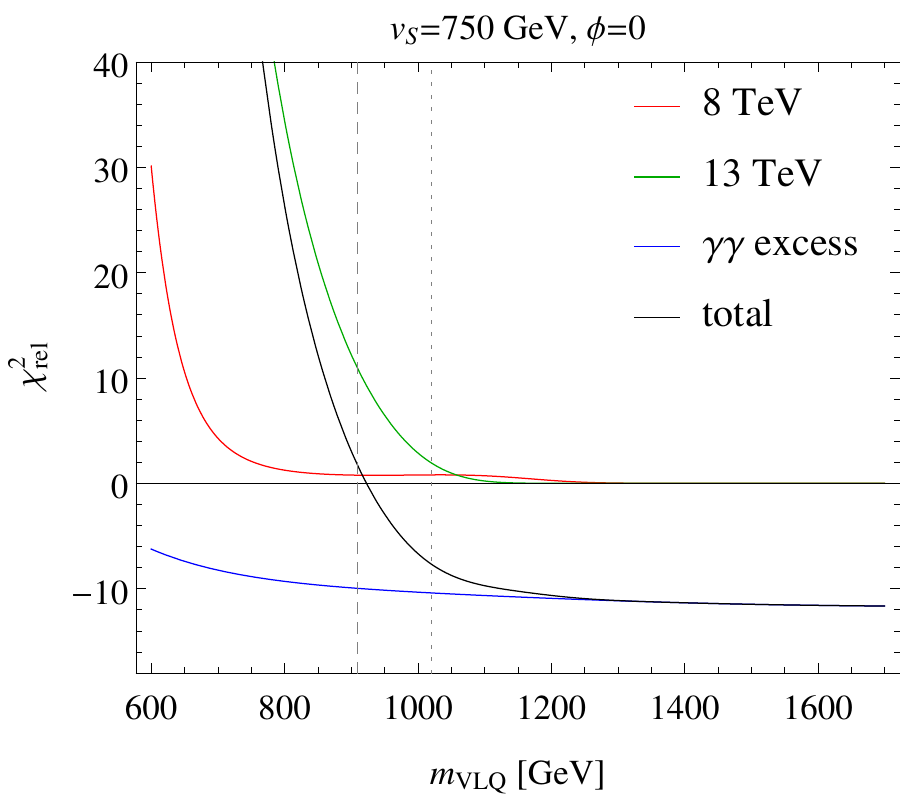}
	\caption{Fit of vector-like quark doublet $(X,T)$ coupling to the first (left) and third (right) generation of SM 
	quarks. The individual contributions to the likelihood from comparison to the $8\, \tev$ and $13\, \tev$ analyses 
	for pair production of VLQs with \toolfont{CheckMATE} and the fit to the diphoton excess are also shown. The dashed (dotted) 
	line marks the limit on the VLQ mass obtained using the CLs method (single sided $95\%$~limit).}
	\label{fig:fits_XT_CM}
\end{figure}

We generate our event samples for vector-like quark pair production with \toolfont{MadGraph} \cite{madgraph} 
and use \toolfont{Pythia6} \cite{Sjostrand:2006za} for parton showering. The cross section is computed at 
NNLO in QCD with \toolfont{Top++} \cite{top++}.
In Fig.~\ref{fig:fits_XT_CM} we compare the likelihood contributions determined 
by \toolfont{CheckMATE} for the $8\, \tev$ and $13\, \tev$ analyses combined with 
the contribution from the diphoton fit. For VLQs coupling to the first 
generation, we find that the fit results are almost indistinguishable
between the different models despite the fact that the branching ratios
to different bosons differ substantially for the $(T,B)$ doublet. The VLQs in the $(T,B)$ doublet decay to $(W,Z,h)$ with branching ratios of approximately $(50\%,\,25\%,\,25\%)$ which roughly agree with the averaged branching ratios of the VLQs in the $(X,T)$ and $(B,Y)$ doublet where one of the VLQs decays exclusively to $W$ bosons and the other decays to $Z$ bosons and $h$ bosons at approximately $50\%$ BR each.
However, in the models with third generation mixing, the different branching ratios to top and bottom 
quarks lead to different signatures at the LHC. Therefore, the results for VLQs coupling to the third generation 
of SM quarks are only valid for the $(X,T)$ VLQ doublet model explored 
here.\footnote{We do not show the LHC results for the other models here since
the $(T,B)$ model is unable to explain the diphoton excess and the $(B,Y)$ model
requires $m_{\text{VLQ}}<600$~GeV which can be clearly seen to be ruled out by
Fig.~\ref{fig:fits_XT_CM}. (Note that the possible presence of $b$-quarks in the final state will actually
tighten the bounds compared to those shown.)}

\begin{table}[htb]
  \begin{tabular}{lccl}
	Experiment & $\sqrt{s}$ [TeV] & \# SRs & Search for \\ \toprule
	ATLAS & $8$ & $15$ & Squarks and gluinos (jets + $E_T^{\text{miss}}$) \cite{ATLAS8squarkGluinos} \\
	ATLAS & $8$ & $1$  & Squarks and gluinos (same-flavour opposite-sign dilepton pair + jets + $E_T^{\text{miss}}$) \cite{ATLAS8dilepton} \\
	CMS   & $8$ & $6$  & Squarks and gluinos (same-flavour opposite-sign dilepton pair + jets + $E_T^{\text{miss}}$) \cite{CMS8dilepton} \\ \midrule
	ATLAS & $13$ & $10$ & Vector-like top quark pairs (1 lepton + jets) \cite{ATLAS13VLQ}\\
	ATLAS & $13$ & $8$ & Gluinos ($>3b$-jets + $E_T^{\text{miss}}$) \cite{ATLAS13gluinos1}\\
	ATLAS & $13$ & $6$ & Gluinos (1 lepton + jets + $E_T^{\text{miss}}$ ) \cite{ATLAS13gluinos2}\\
	ATLAS & $13$ & $7$ & Squarks and gluinos (jets + $E_T^{\text{miss}}$) \cite{ATLAS13squarkGluinos1}\\
	ATLAS & $13$ & $4$ & Squarks and gluinos (2 or 3 leptons + jets + $E_T^{\text{miss}}$) \cite{ATLAS13squarkGluinos2}\\
	ATLAS & $13$ & $1$ & Squarks and gluinos (leptonic-Z + jets  + $E_T^{\text{miss}}$) \cite{ATLAS13dilepton} \\
	CMS   & $13$ & $47$ & Squarks and gluinos (same-flavour opposite-sign dilepton pair + jets + $E_T^{\text{miss}}$)  \cite{CMS13dilepton}\\ \bottomrule
  \end{tabular}
  \caption{List of the analyses with brief descriptions included in the fit of VLQ pair production.}
  \label{tab:listAnalysis}
\end{table}

In a model where the VLQs couple only to the first generation of SM quarks, 
the $8\, \tev$ analyses (mostly the `vanilla' ATLAS jets and $E_T^{\text{miss}}$ 
SUSY search \cite{ATLAS8squarkGluinos}) 
are found to provide the strongest constraints. It is not surprising that
the SUSY search produces the most stringent limit in these cases since the dedicated VLQ
search is designed to only look for top-like partners that decay into the third
generation and consequently produce a number of b-jets in the final state. In
contrast the SUSY search does not require b-jets.

One may wonder why the 8~TeV search gives a stronger limit than the same search does at 13~TeV and this is mainly due to 
the fact that the 13~TeV search has significantly increased both the individual
jet $p_T$ thresholds and the required minimum value of the variable named $m_{\text{eff}}$ (sum of 
jet $p_T$ and $E_T^{\text{miss}}$). For example, the strongest 8~TeV limit 
comes from the 5 jet signal region which requires a hardest jet, $p_T>130$~GeV
and $m_{\text{eff}}>1200$~GeV. In contrast, at 13~TeV these thresholds have been 
increased to $p_T>200$~GeV and $m_{\text{eff}}>1600$~GeV which substantially reduces
the acceptance for our VLQ states.

Using just a single signal region we find a limit using the CLs method of $m_{\text{VLQ}}\gtrsim650$~GeV
on the VLQ states that couple to light quarks. One may find such a result surprising
since one might naively think that VLQ production does not produce a significant
$E_T^{\text{miss}}$ signature. However we find that the neutrinos from $Z\to\nu\nu$ and $W\to\ell\nu$, 
combined with the large production cross-section, do in fact make the supersymmetric searches sensitive to VLQ production.
In addition it is clear that an optimized search for such states would produce significantly
enhanced constraints. In particular we believe that exploiting possible resonant structures in this regard may be
very profitable and these will be explored in a future study\footnote{As stated in the introduction, in the finishing
stages of this study ATLAS released an analysis sensitive to VLQs that included missing energy as a discrimator\cite{ATLAS:2016sno}.}.

Instead of only using a single signal region, we can alternatively combine
all of the orthogonal signal regions in the study and define the VLQ state as being excluded when the fit is
1.64-$\sigma$ worse than the Standard Model (single sided 95\% limit). The combination 
actually leads to a slightly weaker bound of $m_{\text{VLQ}}\gtrsim640$~GeV; the reason for this 
is that the ATLAS on-shell Z + MET excess prefers the production of VLQs near this mass range. However,
we should emphasize that the excess is not statistically significant and we will further discuss this result 
later in this Section.

Nevertheless, when fitting the model to the complete VLQ/SUSY search data set, and including also 
the diphoton signal, we find a large improvement in the fit quality for the entire range of VLQ masses investigated, 
Fig.~\ref{fig:fits_XT_CM}(right). The reason is the 
strength of the diphoton signal and the $\sim3.9\sigma$ excess that is seen over the
Standard Model prediction. Due to the fact that when $\phi=0$ and $v_S=750$~GeV, the $(X,T)$
model predicts a too high diphoton cross-section, we see that the best fit continues 
to improve as we increase the VLQ mass. \\

\begin{figure*}[t]
	\includegraphics[width=0.49\textwidth]{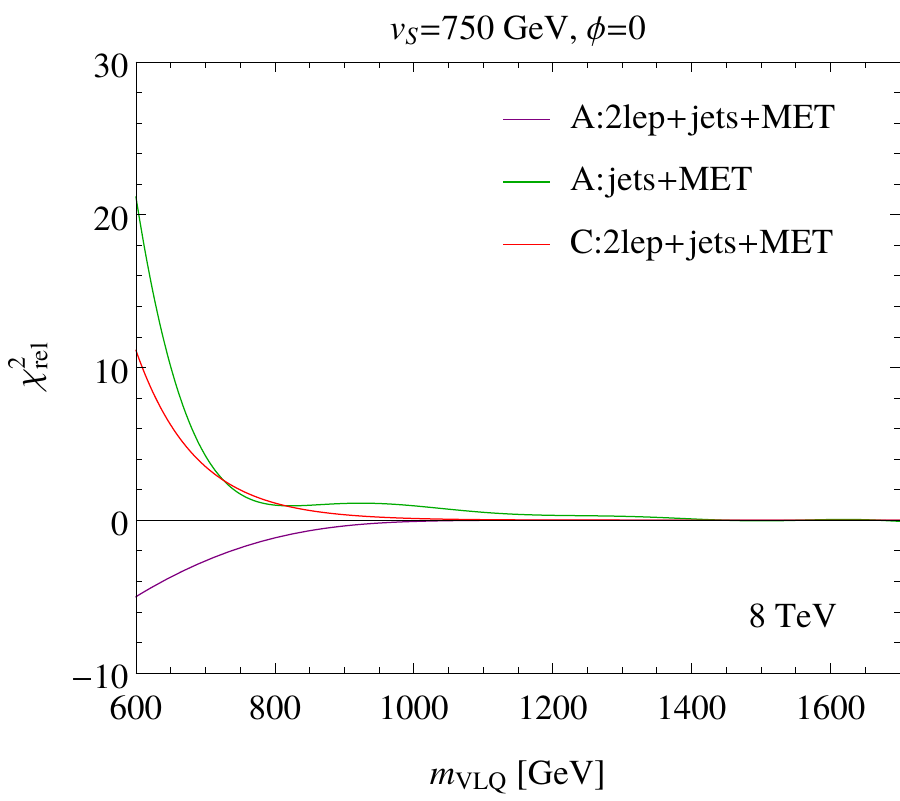}
	\;
	\includegraphics[width=0.49\textwidth]{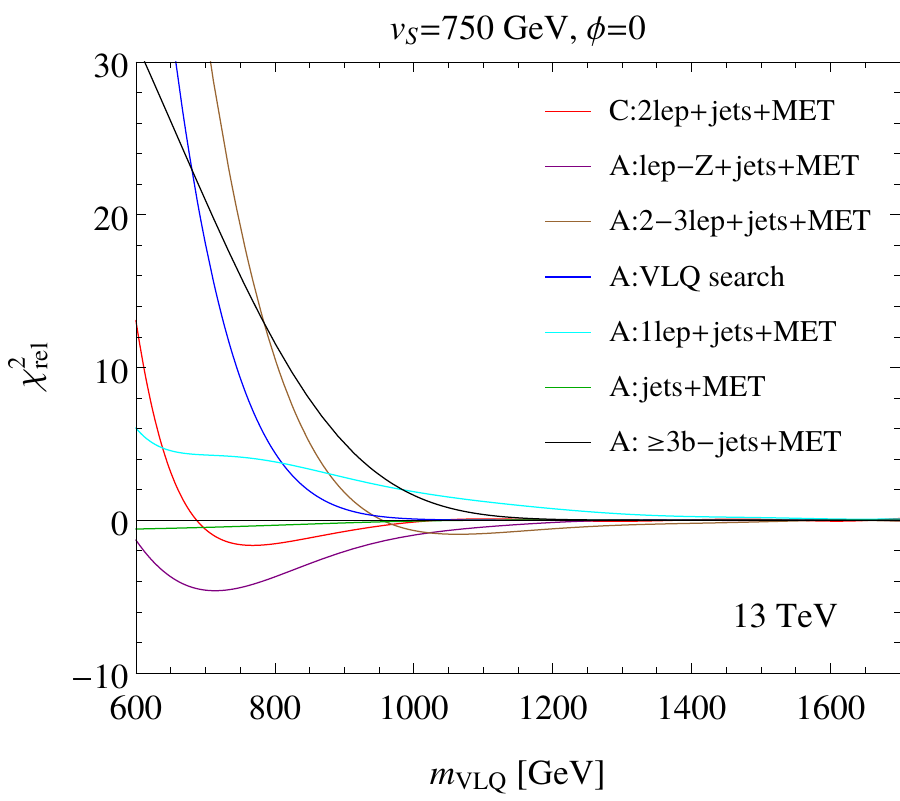}
	\caption{Individual contributions (A=ATLAS and C=CMS) from different LHC searches at $8\,\tev$ (left) and $13\,\tev$ 
	(right) to the relative $\chi^2$ for the $(X,T)$ doublet model coupling to the third generation of SM quarks.}
	\label{fig:XT3_different_analyses}
\end{figure*}

If we now examine the limits in the case where the $(X,T)$ model instead couples to the third generation
we see that the bounds are significantly tighter, Fig.~\ref{fig:fits_XT_CM}(left). The reason
is that both the $X$ and $T$ states will decay via on-shell top quarks (and Higgs Bosons) and these (or the
produced b-quarks) are used by many searches as a way to reduce SM background. 

More surprising is the result that the largest sensitivity and thus the tightest bound on the VLQ states {\it does not} come from 
the dedicated VLQ search \cite{ATLAS13VLQ} which focuses on $T \bar{T}$ production, but instead from 
the ATLAS $13\,\tev$ search for gluinos \cite{ATLAS13gluinos1} with at least 3$b$-jets and $E_T^\text{miss}$, 
Fig. \ref{fig:XT3_different_analyses}. 
For example, the limit at 95\% CLs from this supersymmetric search is $m_{\text{VLQ}}\gtrsim910$~GeV whereas 
the limit from the dedicated VLQ search is only $m_{\text{VLQ}}\gtrsim825$~GeV, so that the supersymmetric search channel
provides a much more stringent constraint!
The difference in these two searches is that the VLQ search is inclusive and only targets final
states with high jet multiplicity (and 1-lepton). In contrast, the supersymmetric search
requires significant $E_T^{\text{miss}}$ as well as $m_{\text{eff}}$ which successfully suppresses the SM background. 

The signal regions of the supersymmetric search which set the tightest limit on this model are
`Gtt-1L-A' and `Gtt-1L-B'. They require 1 lepton in the final state together with 
$E_T^{\text{miss}}~>~200/300$~GeV and $m_{\text{eff}}>1100/900$~GeV respectively. 

The above results show that missing energy from $Z\to\nu\nu$ and $W\to\ell\nu$ that are produced
in the VLQ decays should be much further investigated so as to improve sensitivity to these models.
Conventionally, missing energy has not been considered as important in searches for VLQs but the fact 
that a supersymmetric search that has been optimized for gluinos outperforms the dedicated VLQ
search as found here demonstrates that this is probably misguided.  

From the combination of all orthogonal signal regions we obtain a single sided $95\%$ limit of 
$m_\text{VLQ}\gtrsim 1020\,\gev$. This limit is stronger than the one obtained with the CLs method, 
for two main reasons. Firstly we have combined all analyses together and this leads to a stronger bound
than only considering the best expected limit from a single signal region. Secondly, and more importantly,
the ATLAS $13\,\tev$ search for gluinos with at least $3b$-jets that is most constraining has found  
several under-fluctuations in the relevant signal regions. The commonly used CLs method 
`punishes' the exclusion when such under-fluctuations occur but this is not the case 
in the combined likelihood method that we have used here and thus a much stronger limit is obtained.

\begin{figure*}[t]
	\includegraphics[width=0.49\textwidth]{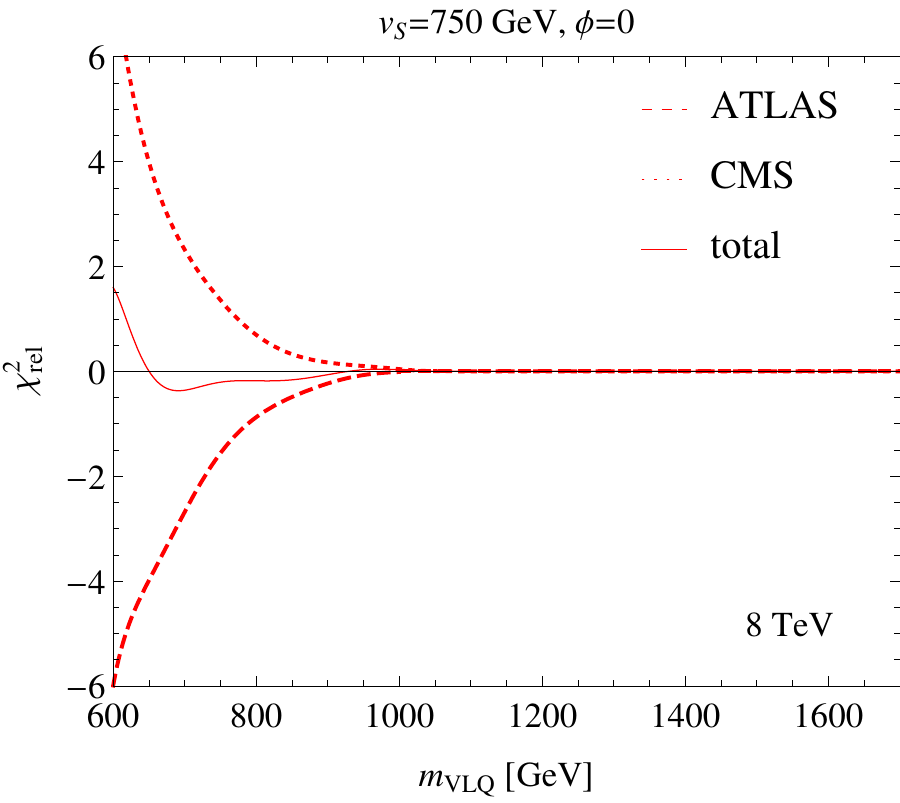}
	\;
	\includegraphics[width=0.49\textwidth]{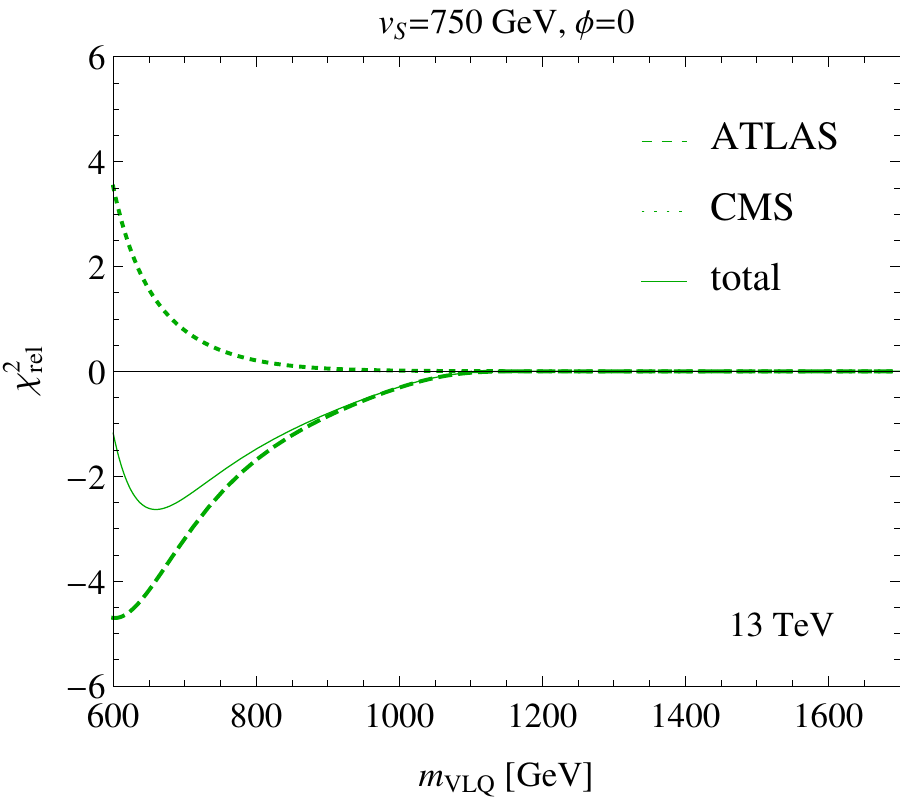}
	\caption{Contributions to the likelihood obtained from comparison of the predictions to the signal region results 
	 sensitive to the ATLAS on-Z excess at $8\, \tev$ (left) and $13\, \tev$ (right) for the $(X,T)$ model.}
	\label{fig:BY_onZ}
\end{figure*}

If we further examine Fig.~\ref{fig:XT3_different_analyses}, we can see that at both 8 and 13~TeV 
the excess that is present in the ATLAS on-shell $Z$+MET gluino search by the negative $\chi^2$
contribution to the fit. Taking into account the other searches we have investigated it
is already clear that any VLQ explanation for this excess does not improve the overall
fit of the model.

Nevertheless, in Fig. \ref{fig:BY_onZ} we now only consider the same ATLAS and CMS
signal regions at 13~TeV and the most similar regions at 8 TeV.
We see that at 8~TeV, the constraint coming from
the non-observation of the excess in CMS almost exactly cancels that from the positive ATLAS result. At
13~TeV, the constraint from CMS is not quite as strong but we still see that the peak significance
is reduced to just 1.5-$\sigma$ and we therefore do not investigate this excess further.

\subsection{Global Fit}

\begin{figure*}[t]
	\includegraphics[width=\textwidth]{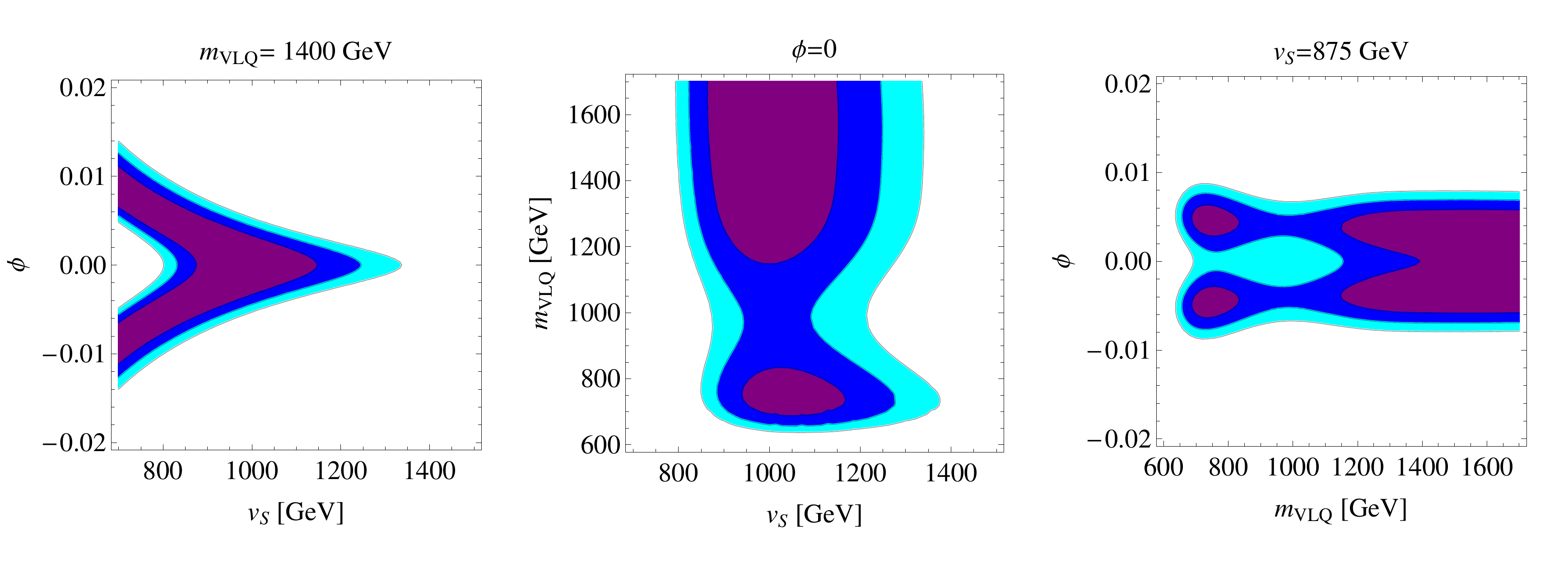}
	\includegraphics[width=\textwidth]{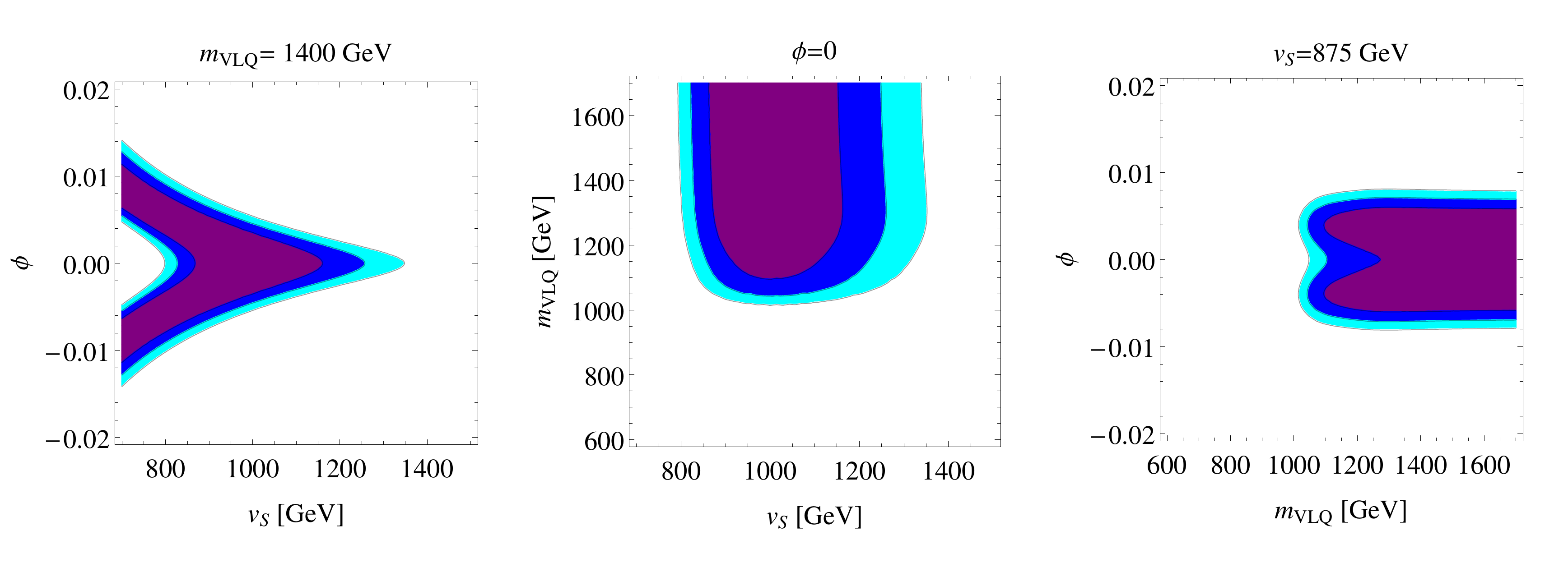}
	\caption{Contours of the likelihood for the $(X,T)$ model for the diphoton contribution taking into account the 
	constraints from VLQ pair production for VLQs coupling to the first (upper) or third (lower) generation of SM 
	fermions. The contours are shown for the $1\sigma$ (purple), $2\sigma$  (dark blue) and $3\sigma$  (light blue) regions.}
	\label{fig:XT_contours_after}
\end{figure*}

In order to understand the model parameters that best fit all of the available data we now
combine our fit of the diphoton signal with the likelihood given by the LHC searches
for VLQ production. Since only the $(X,T)$ model was able to fit the diphoton excess
whilst simultaneously not being excluded by the direct VLQ searches, we now
ignore the other models. In addition, since the decays of the VLQ states 
depend on the generation of the Standard Model quarks  
we now separately discuss the cases where they primarily couple to the first and third generation.

In the upper row of Fig.~\ref{fig:XT_contours_after} we display the case where the VLQs couple 
to the first generation. Comparing with Fig.~\ref{fig:contours} (lower) we see that the
effect of adding the direct production LHC searches is to disfavor models with smaller VLQ masses.
More precisely, models with $m_{\text{VLQ}}\lesssim 630$~GeV are now ruled out at the 3-$\sigma$ 
level. The best fit region is found for large VLQ masses and the main 1-$\sigma$ region
is found for $m_{\text{VLQ}}\gtrsim 1100$~GeV. Large VLQ masses are preferred for two reasons
with the first being that the direct searches for these states disfavor lower masses. Secondly, 
in the $(X,T)$ model with no scalar mixing and $v_S=875$~GeV, the model actually predicts
a too high diphoton rate, and increasing the VLQ mass somewhat reduces the production cross-section.


As explained before, an alternative way to reduce the diphoton rate is to increase the
mixing angle with the SM Higgs to be in the (absolute) range $0.01$--$0.02$. The introduction
of the mixing allows additional decay modes to enter while boosting others and thus produces a better fit. An increase
in $v_S$ also reduces the diphoton rate but is compensated here as the VLQ scalar coupling 
is smaller for fixed $m_{\text{VLQ}}$. A final feature to note
in the best fit contours is that a lighter VLQ region, $700\lesssim m_{\text{VLQ}}\lesssim 850$~GeV,
is compatible with the best fit at 1-$\sigma$. The reason for the existence of this region is that 
the 1.5-$\sigma$ on-shell $Z$+MET ATLAS excess marginally improves the agreement with data here.

Moving on to the case where the $(X,T)$ states couple to the third generation, the effect
of the LHC searches for VLQ states is far more severe. We now see in Fig.~\ref{fig:XT_contours_after} (lower), that models 
with $m_{\text{VLQ}}\lesssim 1050$~GeV are excluded at the 3-$\sigma$ level.
Again we also note that either a non-zero scalar mixing angle or a $v_S \gtrsim 900$~GeV
are required to fit the diphoton excess in the region of VLQ masses studied in this analysis.
\section{Summary}
\label{sec:summary}

In this paper we have examined a model containing an additional scalar singlet and a new vector-like quark 
doublet in order to explore the complementarity of a variety of LHC searches for new states.  
Both the production and decay of the scalar singlet is mediated by the VLQ loop. This scenario is employed as an example
simplified model and could also 
explain the earlier reported diphoton excess observed by the ATLAS and CMS collaborations. 
Furthermore, we employed a number of direct LHC searches for supersymmetric partners and the VLQs themselves. 
Using \toolfont{CheckMATE} we showed how this scenario could be tested and its parameters probed using data from the 
8 and 13 TeV runs of the LHC.

As a very concrete example, we analyzed the models corresponding to the three possible charge assignments 
for the VLQ doublet extension where the VLQ states can also couple to the SM Higgs and the VLQs then 
decay via mixing with their SM partners with the same quantum numbers. Two of these models, with either a 
$(B,Y)$ or $(T,B)$ doublet, predict a diphoton production rate which is too small to explain the apparent 
excess. However, with the $(X,T)$ doublet scenario, we demonstrated that we were able to reproduce correctly 
the size of the apparent excess and impose bounds on the masses and couplings of the VLQ, as well as the 
vacuum expectation value of the singlet scalar and its mixing with SM-like Higgs boson.

Applying the various VLQ and supersymmetric searches, we found that additional constraints are then placed on 
the model to significantly limit the possible parameter space. In the case where the VLQ was assumed to mix predominantly 
to the first generation of SM fermions we found that the 8~TeV supersymmetric search for jets plus missing energy 
provided the strongest constraints and set a lower limit of $m_{\text{VLQ}}\gtrsim650$~GeV. The fact that the 
8~TeV limit was found to be stronger than that derived from the combination of the 13~TeV searches we investigated 
motivates a dedicated 13~TeV effort to constrain such a model.

When the VLQ states mixed predominantly to the third generation SM fermions, we found that the resulting limits
were much stronger requiring that  $m_{\text{VLQ}}\gtrsim910$~GeV. More importantly, however, we found that
the 3-$b$ jet supersymmetric search with missing energy is significantly more sensitive to the production of 
these new VLQ states than was the dedicated VLQ search itself! This result motivates a possible re-appraisal of the 
general VLQ search strategy to examine whether including missing energy as a discriminator can help increase 
sensitivity to these types of models.

In conclusion, we have demonstrated the importance of complementarity of LHC searches for new physics and urge our
experimental colleagues to employ this principle in developing their search analyses.

\subsubsection*{Acknowledgments}

We would like to thank Tim Keller and Jan Sch\"utte-Engel for the implementation of LHC searches into \toolfont{CheckMATE}. JSK thanks Thomas Flacke for discussions. KR is supported by the National Science Centre (Poland) under Grant 2015/19/D/ST2/03136 and the Collaborative Research Center SFB676 of the DFG, ``Particles, Strings, and the Early Universe.''
The work of JSK has been partially supported by the MINECO, Spain, under contract FPA2013-44773-P; Consolider-Ingenio CPAN CSD2007-00042 and  the Spanish MINECO Centro de excelencia Severo Ochoa Program under grant SEV-2012-0249. MK and JT have been supported in part by
the DFG Research Unit 2239 ``New Physics at the Large Hadron
Collider.''  The work of JLH and TGR was supported by the U.S. Department of Energy Office of Science, Contract DE-AC02-76SF00515.

\appendix
\section{Additional decay modes of $h_2$}
\label{sec:appendix}
Here we provide the components of the tree level decays of $h_2$ which are induced by a finite 
mixing ($\phi \neq 0$) with the SM Higgs:
\begin{align}
	\Gamma (h_2 \rightarrow h_1 h_1) &= \frac{s_\phi^2 c_\phi^2 M_{h_2}^3}{32 \pi v_H^2} 
		\left(c_\phi + s_\phi \frac{v_H}{v_S} \right)^2  \left( 1 + 2 u_{h_1} \right)^2 \sqrt{1-4 u_{h_1}}, 
		\quad u_x = \frac{M_x^2}{M_{h_2}^2}, \\
	\Gamma (h_2 \rightarrow t \bar{t}) &= \frac{3 s_\phi^2 M_{h_2} m_t^2}{8 \pi v_H^2} 
		\left( 1 - 4 u_t \right)^2 \sqrt{1-4 u_t},\\
	\Gamma (h_2 \rightarrow ZZ)|_\text{tree} &= \frac{s_\phi^2 G_F^2 v_H^2 M_{h_2}^3}{16 \pi} 
		\left( 1 - 4 u_Z + 12 u_Z^2 \right) \sqrt{1-4 u_Z}, \\
	\Gamma (h_2 \rightarrow WW)|_\text{tree} &= \frac{s_\phi^2 G_F^2 v_H^2 M_{h_2}^3}{8 \pi} 
		\left( 1 - 4 u_W + 12 u_W^2 \right) \sqrt{1-4 u_W}.
\end{align}
	\FloatBarrier
	\bibliography{Z_citations}

\end{document}